\documentclass[twocolumn,a4paper,prb,aps,showpacs]{revtex4}
\usepackage{graphicx}
\usepackage{dcolumn}
\usepackage{bm}

\begin{document}

\title{Magnetocaloric effect in hexacyanochromate Prussian blue analogs}

\author{Esp\'{e}ran\c{c}a Manuel and Marco Evangelisti$^{*}$}
\affiliation{National Research Center on ``nanoStructures and bioSystems at Surfaces'' (S$^{3}$), INFM-CNR,
41100 Modena, Italy}
\author{Marco Affronte}
\affiliation{National Research Center on ``nanoStructures and bioSystems at Surfaces'' (S$^{3}$), INFM-CNR,
and Dipartimento di Fisica, Universit\`{a} di Modena e Reggio Emilia, 41100 Modena, Italy}
\author{Masashi Okubo, Cyrille Train, and Michel Verdaguer}
\affiliation{Chimie Inorganique et Mat\'eriaux Mol\'eculaires, Unit\'e CNRS 7071, Universit\'e Pierre et
Marie Curie, 75252 Paris, France}
\date{\today}

\begin{abstract}
We report on the magnetocaloric properties of two molecule-based hexacyanochromate Prussian blue analogs,
nominally CsNi$^{II}$[Cr$^{III}$(CN)$_{6}]\cdot($H$_{2}$O) and Cr$^{II}_{3}$[Cr$^{III}$(CN)$_{6}$]$_{2}\cdot
12($H$_{2}$O). The former orders ferromagnetically below $T_{C}\simeq 90$~K, whereas the latter is a
ferrimagnet below $T_{C}\simeq 230$~K. For both, we find significantly large magnetic entropy changes $\Delta
S_{m}$ associated to the magnetic phase transitions. Notably, our studies represent the first attempt to look
at molecule-based materials in terms of the magnetocaloric effect for temperatures well above the liquid
helium range.

\end{abstract}


\maketitle

Recent experiments~\cite{tejada00,tejada03,affronte04,evange05a,evange05b} have proven that molecule-based
magnetic materials can manifest a significant magnetocaloric effect (MCE) that, in some
cases,~\cite{evange05a,evange05b} is even larger than for intermetallic and lanthanide-alloys conventionally
studied and employed for cooling applications.~\cite{gschneidner05} This is certain for molecular compounds
based on high-spin clusters with vanishing magnetic anisotropy for which the net cluster spins, resulting
from strong intracluster superexchange interactions, are easily polarized by the applied-field providing
large magnetic entropy changes $\Delta S_{m}$. Their maximum efficiency in terms of MCE takes place, however,
at liquid-helium temperatures. A possible solution to increase the temperature of the maximum $\Delta S_{m}$
is by considering larger anisotropies, although the drawback is that the entropy change gets smaller with
increasing anisotropy.~\cite{zhang01} An alternative route to follow is to strengthen the intermolecular
magnetic correlations which ultimately will give rise to long-range magnetic order (LRMO). The response to
the application or removal of magnetic fields is indeed maximized near the magnetic ordering temperature. In
this respect, extended molecule-based systems like Prussian blue analogs are particularly appealing since
LRMO temperatures up to room temperature have been reported for this class of compounds.~\cite{verdaguer04}

In this Brief Report we report a study of MCE on two Prussian blue analogs, nominally
CsNi$^{II}$[Cr$^{III}$(CN)$_{6}]\cdot($H$_{2}$O) and Cr$^{II}_{3}$[Cr$^{III}$(CN)$_{6}$]$_{2}\cdot
12($H$_{2}$O) (hereafter denoted as NiCr and Cr$_{3}$Cr$_{2}$, respectively), showing different magnetic
ordering at relatively high temperatures. Indeed, NiCr is known to undergo a transition to a long-range {\it
ferromagnetic} ordered state at $\simeq 90$~K,~\cite{gadet92} whereas Cr$_{3}$Cr$_{2}$ behaves as a
long-range ordered {\it ferrimagnet} at temperatures below $\simeq 230$~K.~\cite{mallah99} Both compounds
crystalize in a cubic lattice,~\cite{parameters} the conventional unit cell is depicted in Fig.~1. Exchange
coupling between the two metallic center is ensured by the cyano-bridge. For NiCr, half of the tetrahedral
interstitial sites are occupied by cesium atoms which maintain charge neutrality. Further information on the
structure together with a description of the method of synthesis can be found in Refs.~\cite{gadet92} and
~\cite{mallah99} for NiCr and Cr$_{3}$Cr$_{2}$, respectively. Susceptibility, magnetization and heat capacity
measurements down to 2~K were carried out in a Quantum Design PPMS set-up for the $0<H<7$~T magnetic field
range. All data were collected on powdered samples of the compounds.

\begin{figure}[b!]
\includegraphics[angle=0,width=4.5cm]{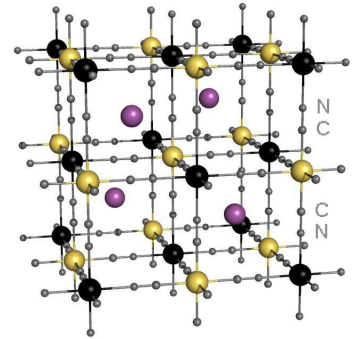}
\caption{(Color online). Sketch of a bimetallic Prussian blue analog. At the vertices, black spheres
represent Cr$^{III}$ ($s=3/2$), whereas lighter-colored spheres represent either Ni$^{II}$ (spin $s=1$) or
Cr$^{II}$ ($s=2$) atoms. Depicted as well are the Cs atoms at the interstitial sites (only relevant for the
NiCr compound).} \label{fig1}
\end{figure}

\begin{figure}[t!]
\includegraphics[angle=0,width=7.2cm]{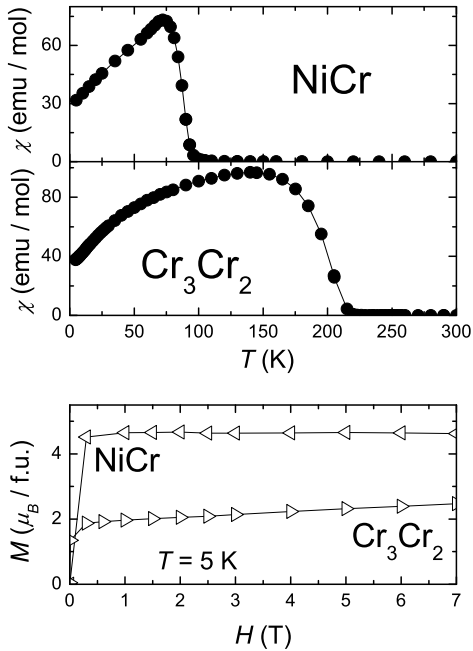}
\caption{Top: Real component of the molar ac-susceptibility $\chi(T)$, collected at $f=1730$~Hz and ac-field
$h_{ac}=10$~G, for NiCr and Cr$_{3}$Cr$_{2}$, as labeled. Bottom: Isothermal magnetization $M(H)$ of both
NiCr and Cr$_{3}$Cr$_{2}$ for $T=5$~K.} \label{fig2}
\end{figure}

For both compounds, Figure~$\ref{fig2}$ shows the real component $\chi(T)$ of the complex susceptibility
collected with an ac-field $h_{ac}=10$~G at $f=1730$~Hz. For NiCr, the abrupt change of $\chi(T)$ at
$T_{C}\simeq 90$~K is ascribed to the transition to a ferromagnetically ordered state, in which
demagnetization effects become important. In fact, the value of 73.5~emu/mol reached by $\chi(T)$ at $T_{C}$
is of the same order as expected for an isotropic NiCr sample at the maximum.~\cite{note1} In the linear
regime of the inverse susceptibility corrected for the demagnetizing field, the fit to the Curie-Weiss law
provides the Curie constant $C=3.2$~emuK/mol and the Weiss constant $\theta=118$~K, in agreement with the
observed ferromagnetic ordering. The constant $C$ is reasonably well accounted for by the expected value of
randomly oriented spins that amounts to 3.1~emuK/mol, assuming $g=2.2$ for Ni$^{II}$ and $g=2$ for Cr$^{III}$
ions. In the ordered phase the saturation magnetization amounts to 4.6~$\mu_{B}$ (Fig.~$\ref{fig2}$), which
corroborates the tendency of the Ni$^{II}$ and Cr$^{III}$ spins to align parallel. For Cr$_{3}$Cr$_{2}$, the
transition to the expected long-range ferrimagnetically ordered state is observed at $T_{C}\approx
230$~K~\cite{note2} (Fig.~$\ref{fig2}$). However, $\chi(T)$ does not reach its maximum at $T_{C}$ but only at
much lower temperatures ($\approx 140$~K), and then it decreases down by lowering $T$. The molar
magnetization of Cr$_{3}$Cr$_{2}$ collected for $T=5$~K increases sharply up to $\approx 1.9~\mu_{B}$ at
$H=0.3$~T and then follows a linear dependence with increasing field, reaching $\approx 2.5~\mu_{B}$ at
$H=7$~T (Fig.~$\ref{fig2}$). As observed by Mallah {\it et al.},~\cite{mallah99} the magnetization fails to
reach the saturation value ($6~\mu_{B}$) expected for a full antiparallel alignment of the Cr$^{II}$ and
Cr$^{III}$ spins. There are two possible reasons for this behavior: (i) Cr$^{II}$ ions can also be in the
low-spin state for which $s=1$, instead of $s=2$; (ii) the magnetic ordered structure is more complex than a
simple ferrimagnet, and spin-canting or reorientation may take place in the ordered state accounting for the
anomalous $\chi(T)$ dependence as well. The latter may be favored by the intrinsic Cr$^{II}$(CN)$_{6}$
vacancies. Likely both reasons coexists, a large fraction of low-spin Cr$^{II}$ would induce structural as
well as magnetic disorder. We refrain from pushing any further the analysis since we do not have enough
experimental information, because our measurements are performed on powder samples.

\begin{figure}[t!]
\includegraphics[angle=0,width=7.2cm]{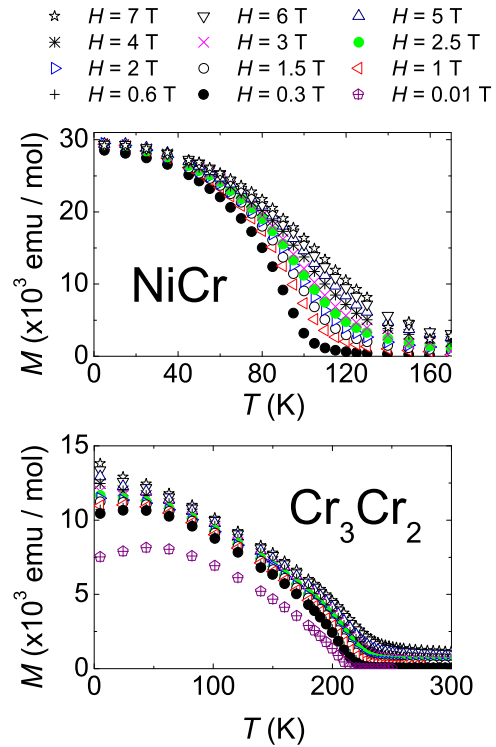}
\caption{(Color online). Field-cooled $M(T)$ curves measured at different applied-fields for NiCr (top) and
Cr$_{3}$Cr$_{2}$ (bottom).} \label{fig3}
\end{figure}

\begin{figure}[h!]
\includegraphics[angle=0,width=7.2cm]{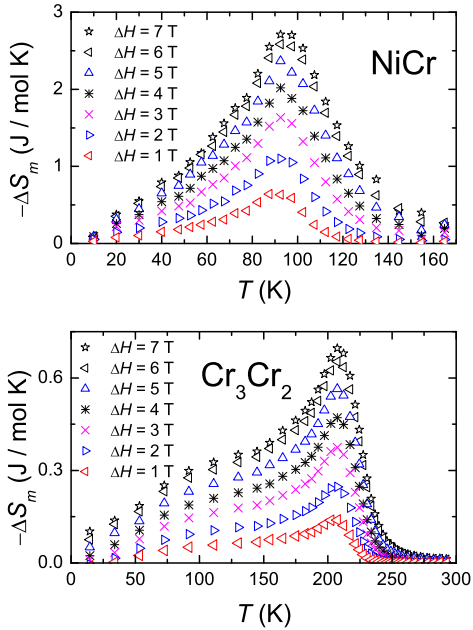} \caption{(Color online). Magnetic entropy change $\Delta S_{m}(T)$
as obtained from $M(T,H)$ data of Fig.~3 for NiCr (top) and Cr$_{3}$Cr$_{2}$ (bottom), for several field
changes $\Delta H$, as labeled.} \label{fig4}
\end{figure}

For a proper evaluation of the MCE of these compounds,~\cite{pecharsky99JAP} we performed systematic
magnetization $M(T,H)$ measurements as a function of temperature and field. Field-cooled $M(T,H)$
measurements for several applied-fields $H$ up to 7~T show spontaneous magnetization below the corresponding
$T_{C}$'s (Fig.~3). In an isothermal process of magnetization, the magnetic entropy change $\Delta S_{m}$ can
be derived from Maxwell relations by integrating over the magnetic field change $\Delta H=H_{f}-H_{i}$, that
is:

\begin{equation}
\Delta S_{m}(T)_{\Delta H}=\int_{H_{i}}^{H_{f}}\frac{\partial M(T,H)}{\partial T}~{\rm d}H.
\end{equation}

From $M(H)$ data of Fig.~$\ref{fig3}$, the obtained $\Delta S_{m}(T)$ for several $\Delta H$
values~\cite{note3} are depicted in Fig.~$\ref{fig4}$. It can be seen that, for both compounds, $\Delta
S_{m}$ has maxima at temperatures corresponding that of the ordering temperatures. Interestingly, $\Delta
S_{m}(T)$ of Cr$_{3}$Cr$_{2}$ keeps relatively large values for a broad $T$-range down to temperatures well
below $T_{C}$. Moreover, we note that $\Delta S_{m}$ increases by increasing $\Delta H$, reaching for $\Delta
H=7$~T the largest values of 2.8~J~mol$^{-1}$K$^{-1}$ for NiCr and 0.70~J~mol$^{-1}$K$^{-1}$ for
Cr$_{3}$Cr$_{2}$, or equivalently 6.6~J~kg$^{-1}$K$^{-1}$ and 0.93~J~kg$^{-1}$K$^{-1}$ for the former and the
latter, respectively.

\begin{figure}[t!]
\includegraphics[angle=0,width=7cm]{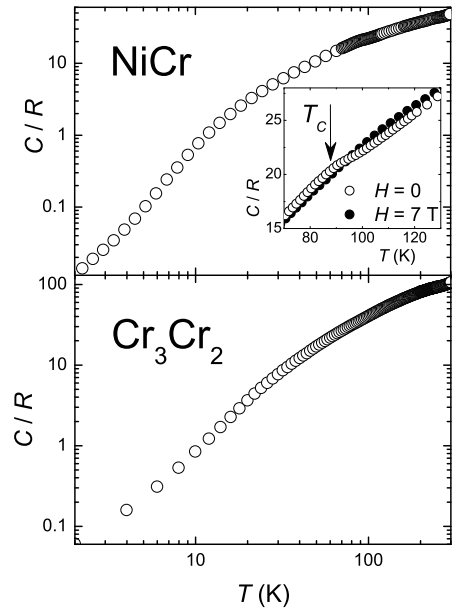}
\caption{Temperature dependence of the specific heat of NiCr (top) and Cr$_{3}$Cr$_{2}$ (bottom) collected
for zero-applied-field. Top inset: Magnification of the region around the ferromagnetic ordering temperature
$T_{C}\simeq 90$~K, for $H=0$ and $H=7$~T.} \label{fig5}
\end{figure}

The magnetocaloric effect of a given magnetic material can be also evaluated from specific heat $C/R$
measurements.~\cite{pecharsky99JAP} The experimental specific heat of NiCr is displayed in Fig.~5 for $H=0$
and $H=7$~T. A careful look at the zero-field curve reveals the onset of the ferromagnetic phase transition
at $T_{C}=90$~K, corroborating the magnetization experiments. It can also be noticed that this contribution
disappears upon application of the magnetic field, proving its magnetic origin. The magnetic contribution is
hardly detectable, the reason is that, at the temperature region of the phase transition, the dominant
contribution arises from thermal vibrations of the lattice.

In order to evaluate MCE from the $C(T,H)$ data of Fig.~5, we first determine the total entropies for $H=0$
and 7~T as functions of $T$, according to

\begin{equation}
S(T)_{H}=\int_{0}^{T}\frac{C(T)_{H}}{T}~{\rm d}T.
\end{equation}

Experimental entropies are obtained integrating down to the lowest achieved $T\approx 2$~K and, obviously,
not from $T=0$~K as in principle required. However, this does not represent an obstacle for our purposes
because at these temperatures the system is fully magnetically ordered and the magnetic entropy is therefore
vanishingly small. For $\Delta H=(7-0)$~T, we then calculate the magnetic entropy change $\Delta S_{m}$ as
well as the adiabatic temperature change $\Delta T_{ad}$,~\cite{pecharsky99JAP} both as function of
temperature. Note that the estimation of the lattice contribution is irrelevant for our calculations, since
we deal with differences between total entropies at different $H$. The results obtained for $\Delta S_{m}$
and $\Delta T_{ad}$ are displayed in Fig.~6. The largest changes are seen near $T_{C}$, where we get $-\Delta
S_{m}=(2.5\pm 0.3)$~J~mol$^{-1}$K$^{-1}$, or equivalently $(5.9\pm 0.7)$~J~kg$^{-1}$K$^{-1}$, and $\Delta
T_{ad}=(1.2\pm 0.1)$~K. It can be noticed that the obtained $\Delta S_{m}$ agrees with the previous estimate
inferred from $M(T,H)$, suggesting that both independent procedures can be effectively used to characterize
the NiCr compound with respect to its magnetocaloric properties.

\begin{figure}[t!]
\includegraphics[angle=0,width=7cm]{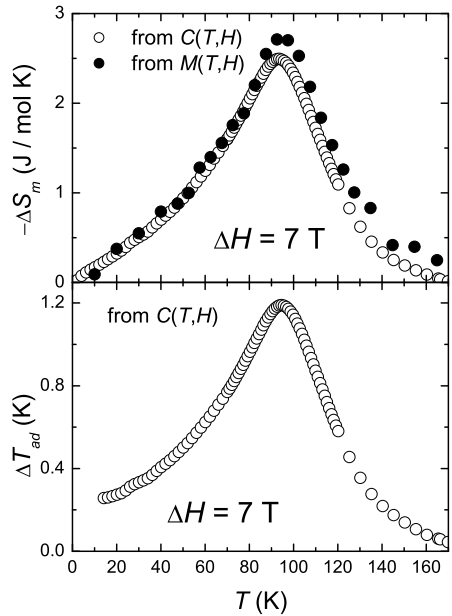}
\caption{For NiCr. Top: $\Delta S_{m}(T)$ as obtained from $C$ (empty dots) and $M$ data (filled dots), both
for $\Delta H=7$~T. Bottom: $\Delta T_{ad}(T)$ as obtained from $C$ data, for $\Delta H=7$~T.}
\label{fig6}
\end{figure}

As for NiCr and for the sake of completeness, we performed heat capacity experiments for Cr$_{3}$Cr$_{2}$ as
well. However, one has to consider that for this system we expect ferrimagnetic order at $T_{C}\simeq 230$~K.
The resulting magnetic moment per formula unit is rather small (Fig.~$\ref{fig2}$), especially taking into
account the large lattice contribution that takes place at these very high temperatures. It is therefore not
surprising that no magnetic anomaly is detectable in the $C(T,H)$ curves. The lower panel of Fig.~5 shows
indeed the measured curve, which essentially does not depend on the applied-field.

Summing up the experimental results here presented, the magnetocaloric response for the molecular materials
NiCr and Cr$_{3}$Cr$_{2}$ shows maximum entropy changes near the ordering temperatures, 90 and 230~K,
respectively. This places the two Prussian blue analogs in a temperature range still below room temperature,
which is dominated, with reference to MCE variations, by lanthanide-based materials.~\cite{pecharsky99} As a
matter of fact, a direct comparison reveals that the NiCr and Cr$_{3}$Cr$_{2}$ compounds have $\Delta S_{m}$
and $\Delta T_{ad}$ that are about an order of magnitude smaller. In spite of such a large gap, our results
are encouraging since they represent a strong improvement with respect to the molecule-based materials
investigated so far in terms of MCE, for which large MCE variations take place only below
10~K.~\cite{tejada00,tejada03,affronte04,evange05a,evange05b} The relevant point is that, as for conventional
materials investigated and proposed for cooling applications,~\cite{gschneidner05} the mechanism of magnetic
ordering can be efficiently exploited also in this class of materials to enhance the magnetocaloric effect.

This work is partially funded by the Italian MIUR under FIRB project No. RBNE01YLKN and by the EC-Network of
Excellence ``MAGMANet'' (No. 515767). E.M. was supported by a fellowship from the EC-Marie Curie network
``QuEMolNa'' (No. MRTN-CT-2003-504880). M.O was supported by Research Fellowships of the Japan Society for
the Promotion of Science for Young Scientists.


\begin{thebibliography}{99}
\bibitem{parameters} For NiCr, the cell parameter is $a=10.50$~\AA~and the molecular weight is
$m_{w}=424.4$~g. For Cr$_{3}$Cr$_{2}$, $a=10.39$~\AA~and $m_{w}=788.3$~g.
\bibitem{note1} An estimate of the demagnetizing factor ($N\approx 0.6$) gives the value of
$\approx 91$~emu/mol for the susceptibility of an isotropic ferromagnetic NiCr sample at the maximum,
assuming a cylindrical approximation of its shape.
\bibitem{note2} The observed ordering temperature of Cr$_{3}$Cr$_{2}$ slightly differs from that
reported in Ref.~\cite{mallah99}. The small discrepancy can be explained by the water content: 12(H$_{2}$O)
here vs. 10(H$_{2}$O) in Ref.~\cite{mallah99}. It is known indeed that the amount of water may (i) vary from
sample to sample, (ii) affect the magnetic properties.~\cite{verdaguer04}
\bibitem{note3} For practical reasons, the measurements at the lowest applied field were carried
out for $H_{i}=10^{-3}$~T, which in our calculations was approximated to zero-applied-field.
\end{thebibliography}
\end{document}